\documentclass[amsmath,amssymb,aps,pra,twocolumn,showpacs,superscriptaddress,nofootinbib]{revtex4}

\usepackage{graphicx,subfigure}
\usepackage{multirow}
\usepackage[colorlinks=true,linkcolor=blue,citecolor=blue,urlcolor=blue]{hyperref}
\usepackage{bbold}
\usepackage{multirow}

\newcommand{\hil}{{\mathcal H}}

\newcommand{\nmax}{{N_{\rm max}}}
\newcommand{\nsites}{{N_{\rm sites}}}

\newcommand{\ev}[1]{{\left\langle #1 \right\rangle}}

\begin{document}

\title{Corner space renormalization method for driven-dissipative 2D correlated systems}

\author{S. Finazzi, A. Le Boit\'{e}, F. Storme, A. Baksic, C. Ciuti}
\email{cristiano.ciuti@univ-paris-diderot.fr}
\affiliation{Laboratoire Mat\'eriaux et Ph\'enom\`enes Quantiques,
Universit\'e Paris Diderot-Paris 7 and CNRS, \\ B\^atiment Condorcet, 10 rue
Alice Domon et L\'eonie Duquet, 75205 Paris Cedex 13, France }
\begin{abstract}
We present a theoretical method to study driven-dissipative correlated systems on lattices with two spatial dimensions (2D). The steady-state density-matrix of the lattice is obtained by solving the master equation in a corner of the Hilbert space. The states spanning the corner space are determined through an iterative procedure, using eigenvectors of the density-matrix of smaller lattice systems, merging in real space two lattices at each iteration and selecting $M$ pairs of states by maximizing their joint probability.  Accuracy of the results is then improved by increasing $M$, the number of states of the corner space,  until convergence is reached.  We demonstrate the efficiency of such an approach by applying it to the driven-dissipative 2D Bose-Hubbard model, describing, e.g., lattices of coupled cavities with quantum optical nonlinearities.
\end{abstract}
\date{\today}
\maketitle


Simulating large quantum systems is a challenging task because their complexity grows exponentially with their size. Indeed, the dimension of the Hilbert space for a multipartite system consisting of $m$ subsystems, each of them described by a space of dimension $N$, is $N^m$. Furthermore, for open systems the physics  can no longer be described only in terms of the eigenstates of an Hamiltonian, requiring instead the knowledge of the density-matrix. In this case, the number of variables to be determined scales as $N^{2m}$, namely the square of the size of the Hilbert space. 

In the last decades, several methods have been proposed to reduce the complexity of this problem. The first attempt in this direction is the renormalization group technique, proposed by Wilson~\cite{wilson} and successfully applied to the Kondo problem. Numerical implementations of this approach are based on the solution of a system with a smaller Hilbert space, where only the relevant physical states with the lowest energies are retained. Ideally, this procedure can be iterated by arbitrarily growing the size of a block system step by step, for instance by doubling the size of the block at each iteration. However, such a numerical implementation of the real-space renormalization group can yield inaccurate results for the system ground state, because the boundary conditions imposed while solving the smaller system might be inappropriate to describe the doubled one~\cite{whiteandnoack}. In the case of one-dimensional systems, a powerful method is represented by the density-matrix renormalization group (DMRG)~\cite{dmrg},  which is based on the selection of the most probable states of the reduced density-matrix of a block, obtained by determining the ground state of the Hamiltonian of a larger section of the lattice. The generalization to two spatial dimensions is challenging and currently under intense study~\cite{Verstraete_review,2D_DMRG}: one approach exploits the artificial description in terms of one-dimensional systems with long-range interactions~\cite{xiang}, while another is based on the generalization of matrix product states~\cite{mps} to projected entangled-pair states~\cite{VertraetePEPS}.  These theoretical methods have been extended to 1D dissipative lattice systems by introducing matrix product density operator algorithms~\cite{VerstraeteMPDO}, time-dependent DMRG~\cite{Daley_review}  and a superoperator renormalization technique~\cite{zwolak}.  These approaches are aimed at solving the master equation governing the dynamics of the density-matrix of the lattice.

Among driven-dissipative systems, lattices of coupled cavity resonators with quantum optical nonlinearities~\cite{Hartmann_Nat_Phys,Greentree,Angelakis} are attracting a considerable interest, e.g. for the realization of non-equilibrum strongly correlated photonic phases~\cite{iacopocristiano}. In particular, the spectacular rise of circuit QED resonators with superconducting Josephson quantum circuits is very promising in this respect both for the realization of strong correlations and for their control~\cite{Houck,SchmidtANNALEN}. So far, several studies have been devoted to non-equilibrium mean-field-like theories~\cite{tomadin,Nissen_PRL,Jin_PRL,alexandre1,Jin_PRA, alexandre2}, based on a Gutzwiller factorization of the density-matrix. Numerical methods beyond mean-field for such systems so far rely on a direct integration of the density-matrix for small size systems\cite{Iacopo2009,FQH} or applications of the matrix product operator techniques mentioned above to one-dimensional cavity arrays~\cite{Hartmann2010,Biella}.  

In this letter, we present a theoretical method to explore the physics of driven-dissipative correlated quantum systems with two spatial dimensions. A corner of the Hilbert space for a lattice system is selected using eigenvectors of the density-matrix solving the master equation for smaller clusters. At each step, two sublattices are merged and $M$  pairs of states are selected to construct a corner basis by maximizing their joint probability.  The degree of accuracy can be controlled by enlarging the number of states of the corner space, until convergence is obtained.  The method is applied to the driven-dissipative 2D Bose-Hubbard model, which describes, e.g., two-dimensional arrays of coupled cavities with quantum nonlinearities.

The general problem we aim to  solve is the Lindblad master equation~\cite{wallsmilburn} for the density matrix $\hat\rho$ of  a driven-dissipative manybody quantum system,
$$
\frac{d\hat\rho}{dt}=\frac{i}{\hbar}[\hat\rho,\hat H]+\sum_j\left[\hat C_j\hat\rho\hat C_j^\dagger
-\frac{1}{2}\left(\hat C_j^\dagger\hat C_j\hat\rho+\hat\rho\hat C_j^\dagger\hat C_j\right)\right],
$$
where $\hat H$ is the Hamiltonian of the lattice system and $\hat C_j$ are operators describing the relaxation of the system due to the interaction with an external bath. We will consider a zero-temperature reservoir for simplicity, although the case of finite temperature can be treated without major complications. To give an example, in the case of a lattice of optical cavities
the jump operators are
$\hat C_j=\sqrt{\gamma_j}\hat b_j$,
where $\hat b_j$ is the photon annihilation operator in the $j$-th cavity and $\gamma_j$ is the corresponding dissipation rate.

The corner space renormalization algorithm we introduce here is based on the following steps: i) determine the steady-state density-matrix for small lattices, for which a direct, brute-force integration of the master equation is possible; ii) merge spatially two pre-determined lattices and select the $M$ most probable product states spanning the so-called corner space; iii) determine the steady-state solution of the density-matrix in the corner space; iv) increase the dimension $M$ of the corner until convergence of the observables is achieved; v) in order to create a larger lattice, go back to step ii). 

\begin{figure}
\centerline{
\includegraphics[width= 9.5 cm]
{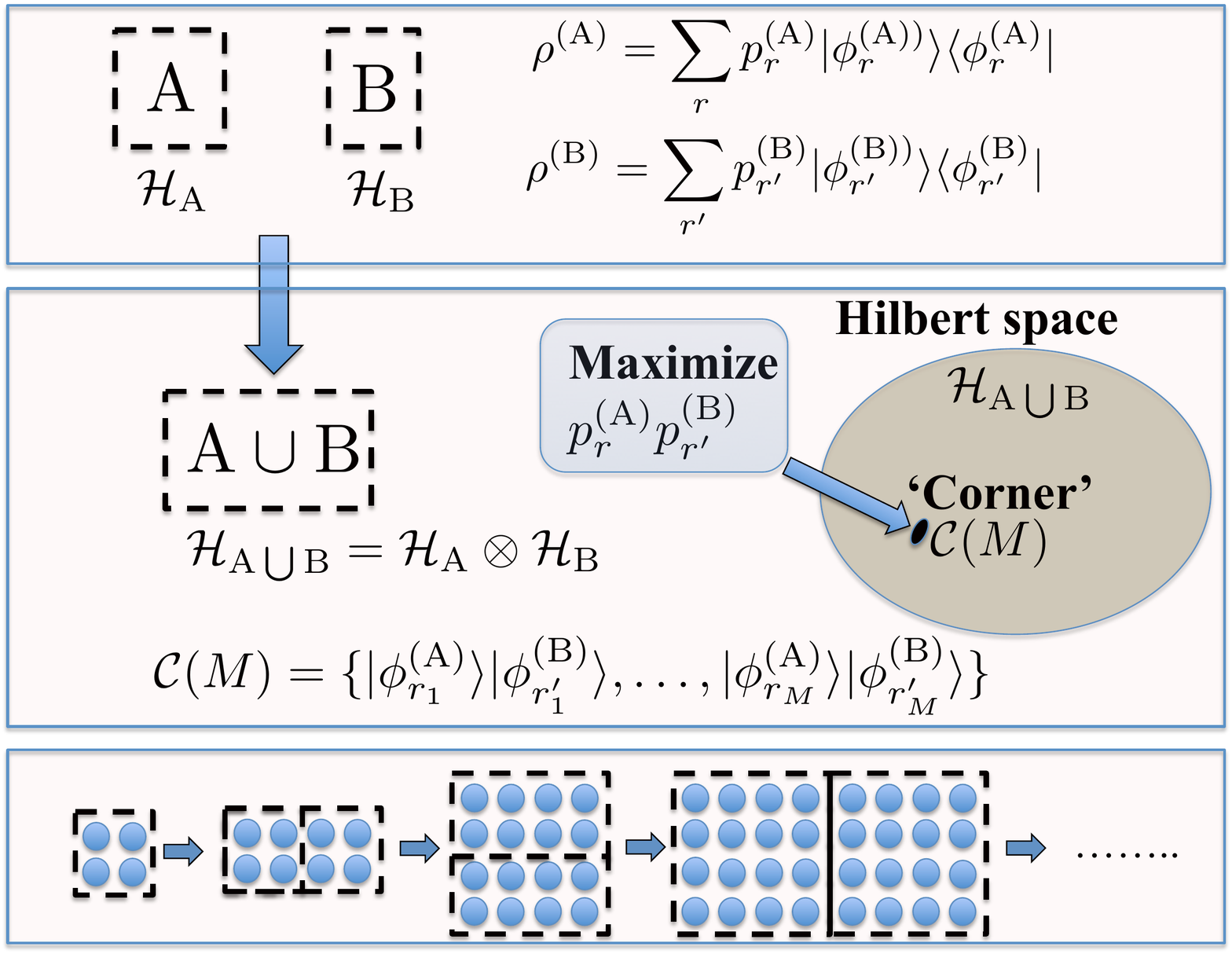}
}
\caption{\label{sketch} Sketch of the corner space renormalization method. }
\end{figure}

Here, we describe in detail the crucial steps iii) and iv), i.e., the selection of the corner space. As sketched in Fig. \ref{sketch}, let us suppose that we know the  solution for the steady-state density matrices $\rho^{(\rm A)}$ and $\rho^{(\rm B)}$ for two lattices ${\rm A}$ and ${\rm B}$. 
If we want to consider a lattice obtained by merging spatially the two lattices ${\rm A}$ and ${\rm B}$, the corresponding Hilbert space is $ \hil_{({\rm A} \cup {\rm B})} = \hil_{\rm A}\otimes\hil_{\rm B}$  where  $\hil_{{\rm A}}$ and $\hil_{\rm B}$ are the Hilbert spaces of  ${\rm A}$ and ${\rm B}$. 
Each density-matrix operator can be diagonalized as $
\rho^{(\rm A)} = \sum_{r} p_{r}^{(\rm A)}  \vert \phi_{r}^{(\rm A)} \rangle  \langle \phi_{r}^{(\rm A)} \vert$, where the states $\vert \phi_{r}^{(\rm A)} \rangle$ form an orthonormal basis for $\hil_{\rm A}$ and $p_{r}^{(\rm A)}$ are the corresponding probabilities. Analogous notations hold for the system  ${\rm B}$. 
Each ket $\vert \phi_{r}^{(\rm A)} \rangle$ represents a manybody state which can have strong correlations within the system $A$.
To select a small `corner'  $\mathcal C(M)$ of the larger space $ \hil_{{\rm A} \cup {\rm B}}$, we will consider the subspace spanned by the $M$ most probable states of the form $\vert \phi_{r}^{(\rm A)} \rangle  \vert \phi_{r'}^{(\rm B)}\rangle$ ranked according to the {\it joint} probability $p_{r}^{(\rm A)} p_{r'}^{(\rm B)}$. Let us call $\vert \phi_{r_1}^{(\rm A)} \rangle \vert \phi_{r_1'}^{(\rm B)}\rangle$  the most probable product state, i.e. such that $p_{r_1}^{(\rm A)} p_{r_1'}^{(\rm B)} \geq p_{r}^{(\rm A)} p_{r'}^{(\rm B)}$ for every value of $r$ and $r'$. We will call $\vert \phi_{r_2}^{(\rm A)} \rangle \vert \phi_{r_2}^{(\rm B)}\rangle$ the second most probable product state and so on so forth. In other words, we will consider the subspace generated by the orthonormal basis $\{ \vert \phi_{r_1}^{(\rm A)} \rangle \vert \phi_{r'_1}^{(\rm B)}\rangle, \vert \phi_{r_2}^{(\rm A)} \rangle  \vert \phi_{r'_2}^{(\rm B)}\rangle , ..., \vert \phi_{r_M}^{(\rm A)} \rangle  \vert \phi_{r'_M}^{(\rm B)}\rangle \} $,
where $p_{r_1}^{(\rm A)} p_{r'_1}^{(\rm B)} \geq p_{r_2}^{(\rm A)} p_{r'_2}^{(\rm B)} \geq ... \geq p_{r_M}^{(\rm A)} p_{r'_M}^{(\rm B)}$, i.e. we select the $M$ most probable pairs of states\footnote{Note that  $p_{r_j}^{(\rm A)}$ can be smaller than $p_{r_{j+1}}^{(\rm A)}$ or $p_{r'_j}^{(\rm B)}$ can be smaller than $p_{r'_{j+1}}^{(\rm B)}$, but always
$p_{r_j}^{(\rm A)}  p_{r'_j}^{(\rm B)} \geq p_{r_{j+1}}^{(\rm A)} p_{r'_{j+1}}^{(\rm B)}$.}.
Note that a generic state belonging to the corner space, namely of the form  $\vert \Psi \rangle  = \sum^{M}_{s=1} c_{s} \vert \phi_{r_s}^{(\rm A)} \rangle  \vert \phi_{r'_s}^{(\rm B)}\rangle$, can describe strong correlations and quantum entanglement between systems $\rm A$ and $\rm B$ while keeping correlations within $\rm A$ and $\rm B$.
We emphasize that by increasing arbitrarily the number $M$ of states  in the corner space, the method becomes exact, because the considered basis spans the entire Hilbert space. Of course, the method is useful only when the number of states $M$ required to reach convergence is small enough to be treated numerically. This ultimately depends on the degree of correlation of the considered system. 

\begin{table}[t!]
\begin{tabular}{ccccc}
\hline\hline
$M$	&	$n$	& $\Re (\langle b \rangle) $ & $g^{(2)}_{<j,l>}$
\\
\hline
20	&	0.09443		& 0.2772 	 & 1.029  \\
50	&	0.09469		& 0.2770 	 & 0.9693  \\
100	&	0.09513		& 0.2768 	 & 0.9652 \\
 200	&  	0.09541  		& 0.2767  &  1.061 \\
400	&	0.09544		& 0.2767 	 & 1.058\\
800 & 0.09549(3)&  0.27671(5) & 1.0644(1)\\
1600 & 0.09547(3) & 0.27672(6) & 1.0643(1)\\
{\it 65536} & {\it 0.0954(1)} & {\it 0.2764(2)} & {\it 1.0643(3)}\\
\hline\hline

\end{tabular}
\caption{Corner method results for the driven-dissipative Bose-Hubbard model with periodic boundary conditions and the following parameters: $4 \times 4$ square lattice ($z = 4$), $U =  + \infty$ ($N_{max} = 1$, hard-core bosons), $J /\gamma = 1$, $F/ \gamma = 2$, $\Delta \omega/ \gamma = 5$.
The numbers in parenthesis indicate the statistical errors on the last significative digit due to  finite Monte Carlo sampling when applied.
In this example, the  dimension of the full Hilbert space is $2^{16} = 65536$. The case of $65536$ states has been solved by an independent Monte Carlo wavefunction code using a Fock basis for the entire space and sparse matrix calculations.  
\label{hardcore_benchmark} }
\end{table}

\vspace{0.5cm}
\begin{table}[t!]
\hspace{0.2em}
\begin{tabular}{ccccc}
\hline\hline
$M$	&	$n$	& $\Re (\langle b \rangle) $ &  $g_{2}$ & $g^{(2)}_{<j,l>}$ \\
\hline
20	&	0.0902		& 0.1967	&	1.646		& 1.28\\
50	&	0.1006		& 0.1907	&	1.513		& 1.34\\
100	&	0.1044		& 0.1886	&	1.454		& 1.26\\
200	&	0.0968		& 0.1922	&	1.324		& 1.51\\
400	&	0.1006		& 0.1905	     &	 1.291		& 1.51\\
800	&	0.1009(2)  	& 0.1903(3)   &  1.242(3)           & 1.57(2) \\
1600 &     0.1014(2)          & 0.1896(2)   & 1.226(3)            & 1.58(2) \\
3200 &     0.1002(2)          & 0.1897(2)   &  1.185(2)           & 1.63(2) \\
6400 &     0.0994(2)          &  0.1899(2)  & 1.179(3)            & 1.63(1) \\
\hline\hline
\end{tabular}
\caption{
Parameters: $4 \times 4$ square lattice with periodic boundary conditions, $U/\gamma =  20$, $J /\gamma = 3$, $F/ \gamma = 2$, $\Delta \omega/ \gamma = 5$.
A number $N_{max} = 3$ of bosons per site has been considered. In this case, the dimension of the full Hilbert space is $4^{16}  \simeq 4.3 \cdot 10^9$. 
  \label{tab:method}}
\end{table}

Concerning step iii), namely the determination of the steady-state density-matrix, it is worth pointing out that when the number $M$ of states in the corner space is small enough, the master equation can be solved  by direct numerical integration in time ($M$ typically up to a few hundreds). For larger values of $M$ (up to a number of the order of $10^4$ depending on the sparsity of the Hamiltonian matrices), a more efficient method is based on a stochastic technique\cite{PlenioRMP}, the so-called Monte Carlo wavefunction algorithm \cite{Dalibard,mwf,Carmichael1993}.  Such algorithm computes the density-matrix of the system by averaging over quantum trajectories of the wavefunction in presence of random quantum jumps. 

As a first illustration of the corner space renormalization method, we show results for the driven-dissipative Bose-Hubbard model in 2D square lattices. The corresponding Hamiltonian ($\hbar = 1$)  in the frame rotating at the pump frequency and in the case of homogeneous pumping reads:
%
 $$\hat H =\sum_{j} ( -\Delta\omega\,\hat b_j^\dagger \hat b_j 
 	+ \frac{U}{2}\,\hat b_j^\dagger \hat b_j^\dagger\hat b_j\hat b_j 
	+ F  (\hat b^\dagger_j + \hat b_j))  
	-\frac{J}{z}\sum_{<j,l> }   \hat b_j^\dagger b_l , $$
%
where $\Delta\omega=\omega_p-\omega_c$ is the detuning between the pump and the bare boson frequency, $U$ is the  on-site boson-boson interaction and $F$ is the pump field. $J$ is the hopping coupling, $z$ is the coordination number and  $\sum_{<i,j>}$ denotes the sum over all the couples of nearest neighbors. For simplicity, we have fixed the phase of the pump in such a way that $F$ is real.  Finally, each site is subject to losses with a dissipation rate $\gamma$.

\begin{figure}
\centerline{
\includegraphics{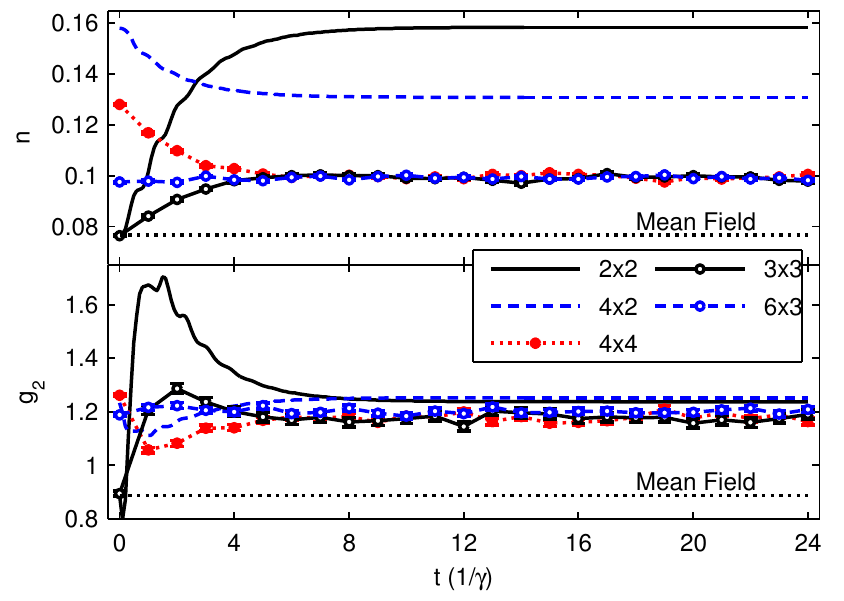}}
\caption{\label{fig:timeevol} Evolution of $n$ and $g_2$ versus  time $t$ (units of $1/\gamma$) for the driven-dissipative Bose-Hubbard model with periodic boundary conditions on lattices of various size for the following parameters: $U/\gamma = 20$, $J/\gamma = 3$, $F/\gamma = 2$, $\Delta \omega/\gamma = 5$ . Solid lines represents evolutions performed by direct integration of the master equation, while points depict  Monte Carlo wavefunction calculations. When error bars are not shown, the statistical error is smaller than the point size.  The black-dotted lines represent the mean-field values. The initial conditions are explained in the text.
}
\end{figure}

\begin{figure}
\centerline{
\includegraphics{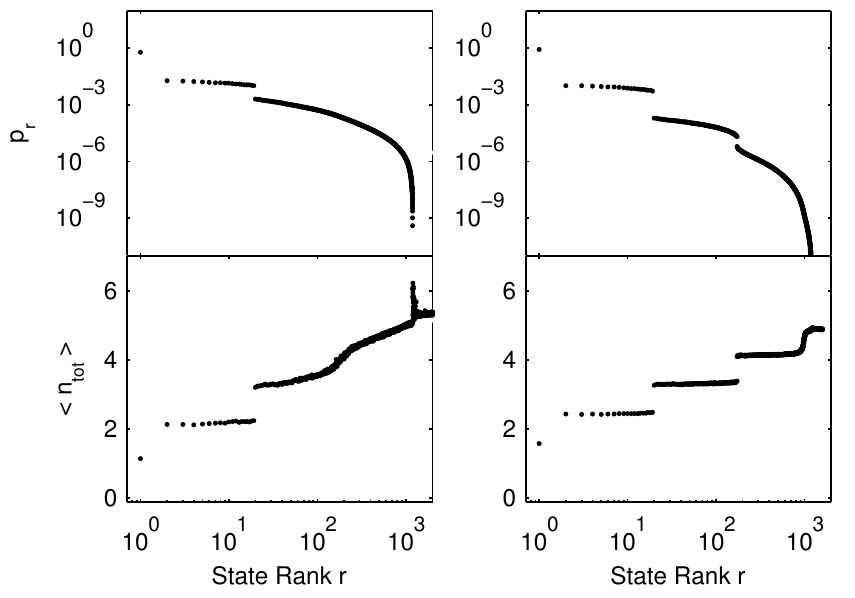}}
\caption{Probabilities $p_r$ (top panels, logarithmic scale) and expectation value of the total boson population $ \langle n_{\rm tot} \rangle = \sum_j \langle n_j \rangle$ for the orthonormal eigenvectors $\vert \Psi_r \rangle$ of the steady-state density-matrix ($\hat{\rho} = \sum_{r}\ p_r \vert \Psi_r \rangle \langle \Psi_r \vert $ and $p_r \geq p_{r+1}$). The state rank $r$ is in logarithmic scale. Lattice size: $6\times3$. Driving parameters: $F/\gamma = 2$, $\Delta \omega/\gamma = 5$. Left: $U/\gamma = 20$ and $J/\gamma = 3$. Right: hardcore bosons with $J/\gamma = 1$.\label{proba}}
\end{figure}

In the following, we will consider the case of periodic boundary conditions. In Table \ref{hardcore_benchmark} we show results for a $4 \times 4$ square lattice of hard-core bosons ($U = + \infty$), i.e. for which the maximum number of photons per site to be considered is $N_{max} = 1$. These results have been obtained starting from a $2 \times 2$ lattice for which a brute-force determination of the steady-state solution of the master equation is possible.  Merging two $2 \times 2$ lattices, we get results for a $4 \times 2$ lattice and, repeating the doubling procedure, for the $4 \times 4$ case. The dimension of the full Hilbert space for hard-core bosons on a $4 \times 4 $ lattice is $2^{16} = 65536$. Although very heavy, the master equation resolution in the full Hilbert space has been performed by an independent Monte Carlo wavefunction code using a Fock basis for the entire space and sparse matrix calculations. This way, we have been able to benchmark the results obtained with a small corner to the exact results. 
The table reports results for the boson population per site $n = \langle \hat{b}_j^\dagger \hat{b}_j \rangle$,  the real part $\{ \Re (\langle b_j \rangle) \}$ of the bosonic coherence and the nearest-neighbor correlation
$ g^{(2)}_{\langle j,l \rangle}= \frac{\ev{\hat b_j^\dagger\hat b_l^\dagger\hat b_j\hat b_l}}{n_j n_l}$. For hard-core bosons, the on-site second-order correlation function $g_2= \frac{\ev{\hat b_j^\dagger \hat b_j^\dagger \hat b_j \hat b_j}}{n_i^2}$ is trivially equal to $0$ since two bosons are not allowed to be on the same site.
 Note that $g^{(2)}_{\langle j,l \rangle}= 1$ for a factorized Gutzwiller-like density-matrix $\hat{\rho}_{\rm G} = \bigotimes_j \hat{\rho}_j$ where $\hat{\rho}_j$ is the reduced density-matrix of the $j$-th site. The mean-field approach is equivalent to taking a self-consistent Gutzwiller density-matrix, assuming all the sites identical. Hence  the difference $( g^{(2)}_{<j,l>}-1)$ quantifies the degree of correlations beyond mean-field between nearest neighbors. Remarkably, for the parameters given in the caption of Table \ref{hardcore_benchmark}, we get a very accurate result for a small number $M = 200$ (negligibible error for $n$, $0.1 \%$ for the bosonic coherence and  $0.3 \%$ for $g^{(2)}_{\langle j,l \rangle}$). In Table \ref{tab:method}, we show results for soft-core bosons  with a larger hopping coupling ($U/\gamma = 20$, $J/\gamma = 3$) and a cut-off number $N_{max} = 3$ of bosons per site (we have verified that this is the cut-off number per site required to get convergence).   A cut-off $N_{max} = 3$ for a $4 \times 4$ lattice implies a Hilbert space dimension equal to $4^{16} \simeq 4.3 \cdot 10^{9}$. As shown by the convergence progression in Table \ref{tab:method},  results with deviations below $1\%$ are reached already for a corner space dimension $M = 3200$, hence six orders of magnitude smaller than the full Hilbert space for a system exhibiting large correlations ($g^{(2)}_{\langle j,l \rangle} - 1 = 0.63$). 

An example of the temporal dynamics leading to steady-state solutions is reported in Fig. \ref{fig:timeevol}, plotting $n$ and $g_2$ for different lattice sizes.
 The corner method results are compared with the non-equilibrium mean-field approach used in Refs. \cite{alexandre1,alexandre2}, based on the exact analytical solution of the master equation for the one-site problem\cite{drummondwalls}.  
  The initial condition for the density-matrix dynamics for the $2 \times 2$ lattice is the mean-field solution. After a transient, a steady-state solution is obtained. The initial condition for the $4 \times 2$ lattice is constructed from the steady-state solution of the $2 \times 2$ lattice and so-on so forth. We have also merged  $3 \times 1$ clusters to get the $3 \times 3$ lattice and then the $6 \times 3$ case by doubling.
   We see that the steady-state observables for the  $3\times 3$, $4\times 4$ and $6 \times 3$ lattices with periodic boundary conditions tend to converge to the same value, so the results are already approaching those for a lattice with an infinite number of sites.  The finite spatial range of the correlations of the driven-dissipative system is responsible for such relatively quick convergence. For the parameters in Fig. \ref{fig:timeevol}, the deviations from the mean-field theory are around $20 \%$ for $n$  and $g_2$.  Since the driving is homogeneous and the considered boundary conditions are periodic,  shortcomings due to conflicting boundary conditions in the doubling procedure do not apply here. 
\begin{table}[t!]
\begin{tabular}{ccccccc}
\hline\hline
\multicolumn{1}{c}{} &
\multicolumn{2}{|c|}{Mean-field} &
\multicolumn{4}{|c}{Corner method}
\\
\hline
 $U /\gamma$  &  
 $n$  & $g_2$ &
 $\nsites^{(M)}$&  
 $n$  & $g_2$  & $g^{(2)}_{\langle j,l\rangle}$
   \\
 \hline\hline
  \multirow{2}{*}{$\infty$ }& \multirow{2}{*}{0.0953} & \multirow{2}{*}{0}  &
  $8\times4^{(1600)}$ &   0.09527(2) & 0 & 1.0436(3)   \\
  & & &
 $8\times8^{(8000)}$ &    0.0948(2) & 0 & 1.0237(6)
 \\
 \hline 
 \multirow{2}{*}{20}&  \multirow{2}{*}{0.125} & \multirow{2}{*}{0.836} &
  $4\times4^{(3200)}$&   0.1281(4)& 0.859(4) &1.172(5)
  \\
 & &  &
  $6\times3^{(6400)}$ &    0.1282(9) & 0.858(9) & 1.173(4)
 \\
 \hline
 
 \multirow{2}{*}{$20^{*}$}&  \multirow{2}{*}{0.0768} & \multirow{2}{*}{0.8879}  &
 
 $4\times4^{(6400)}$&   0.0994(2) & 1.179(3)&1.63(1)\\
 & &  &
 
 $6\times3^{(6400)}$ &   
 0.0992(1) & 1.202(4) & 1.65(1)
 \\
 \hline

 \multirow{2}{*}{10}& \multirow{2}{*}{0.9587} & \multirow{2}{*}{0.6088}  &

 $4\times2^{(6400)}$ &   
 0.9275(8) & 0.631(1) & 1.0127(8) \\
 & &  &
 
 $3\times3^{(8000)}$&  
 {  0.9281(9) } & {  0.617(1) } & { 1.0069 (6)}
 \\
 \hline
 \multirow{1}{*}{1}&  \multirow{1}{*}{0.1156} & \multirow{1}{*}{1.265}  &

 $16 \times 8^{(600)}$ &  0.1156 & 1.259 & 0.9897 
\\
\hline
\multirow{1}{*}{0.5}&  \multirow{1}{*}{0.1126} & \multirow{1}{*}{1.112}  &

 $16 \times 16^{(400)}$ &  0.1126  
  & 1.1105  & 0.9941 
\\
\hline
 %

\hline\hline
\end{tabular} 
\caption{Steady-state expectation values for lattices (periodic boundary conditions) with different sizes, calculated via the Gutzwiller mean-field theory and  the corner space renormalization method. $M$ is the dimension of the corner space. Parameters: $J/\gamma = 1$ (except the third line with the $^*$ sign, obtained with $J/\gamma = 3$), $F/\gamma = 2$ and $\Delta \omega/\gamma = 5$. The  maximum number of bosons per site is $\nmax = 1$ for hardcore bosons, $\nmax = 3$ for $U/\gamma = 20$,  $\nmax = 5$ for $U/\gamma = 10$, $\nmax = 4$ for $U/\gamma = 1$ and $0.5$.
 \label{table_Bose_Hubbard}}
\end{table}

It is insightful to look at the diagonal decomposition of the calculated density-matrix, namely $\hat{\rho} = \sum_{r=1}^M \ p_r \vert \Psi_r \rangle \langle \Psi_r \vert $ where  $p_{r} \geq p_{r+1}$. In Fig. \ref{proba}, we show an example of the probability distribution $p_r$ (top panels, logarithmic scale) and the expectation value of the total number of bosons in the lattice (bottom panels) versus the state rank $r$ for a $6 \times 3$ lattice of  soft-core bosons with $U = 20 \gamma$ (left panels) and hard-core bosons (right panels).
In both cases, the probability drops sharply by several orders of magnitudes when the rank $r$ is large enough, confirming the achieved convergence of the corner space dimension. In the hard-core boson case, a rather well definite shell structure is apparent. The first state ($r = 1$), which captures a large part of probability, is followed by shells of states having close probabilities and densities. In the case of a homogeneous system, a factorized Gutzwiller density-matrix with each site having the same reduced-density matrix leads to a shell structure with exactly flat plateaux structures due to symmetry reasons. In fact, a permutation of the role of the different sites does not change the probability of a state and observables like $n_{tot}$, which is a sum over all the sites. In the right panel of  Fig. \ref{proba}  (hard-core boson case), the situation is qualitatively close to the Gutzwiller case, even though the plateaux are not exactly flat. In the case of soft-core bosons in the left panel of Fig. \ref{proba}, a first shell is clearly visible, while higher shells merge into a continuous curve where the different quantities increase gradually, denoting a large degree of correlations (indeed $g^{(2)}_{<j,l>} - 1  \simeq 0.6$ in the case considered). 

In Table \ref{table_Bose_Hubbard}, we summarize  illustrative results for different lattice sizes   with periodic boundary conditions and  compare them to the Gutzwiller mean-field solutions\cite{alexandre1,alexandre2}, using the same excitation parameters ($F/\gamma = 2$ and $\Delta \omega/\gamma = 5$). The convergence of the results with increasing corner dimension $M$ has been checked as well as the required  maximum number $N_{max}$ of bosons per site. It is apparent that in the considered case the mean-field theory gives rather  good results for hard-core bosons and a large $8 \times 8$ lattice, as quantified by a $g^{(2)}_{\langle i,j \rangle} - 1 \simeq 0.02$. Significant deviations are instead present when the on-site interaction $U$ is competing with the hopping coupling $J$ (the cases with $U/\gamma = 20$ and $J/\gamma =1$ and $3$ in Table \ref{table_Bose_Hubbard}). The value
 for $U/\gamma = 10$ and $J/\gamma = 1$ is close to a two-photon resonance \cite{alexandre2} and indeed the the population of bosons per site is much higher (close to one boson per site) with the on-site $g_2$ correlation function quite close to $0.5$. 
 For $U/\gamma = 0.5$, it is possible to simulate very large lattices (a $16 \times 16$ lattice is reported) with a very small number of states ($M = 400$). 
      
In conclusion, we have presented a theoretical method for driven-dissipative 2D correlated lattice systems. The proposed numerical algorithm follows a hybrid real-space renormalization group approach where the states are selected on the basis of joint probabilities.  We have successfully demonstrated the efficiency of such a method by applying it to the driven-dissipative Bose-Hubbard model on 2D square lattices.  Unlike mean-field theories, where the decoupling approximation is not controlled, the present numerical method allows us to get results with controllable accuracy. The method has therefore the potential to become a precious tool to benchmark analytical theories and study strongly correlated open systems with more than one spatial dimension.   Future studies will explore the physics of  2D arrays of nonlinear cavities with complex elementary cells (including disorder), geometric and spin frustration as well as the role of artificial gauge fields in extended lattices. 

C. C. acknowledges support from ERC (via the Consolidator Grant 'CORPHO' No.
616233), from ANR (projects QPOL and QUANDYDE) and from Institut Universitaire de France.


\newpage
\end{document}